\documentclass[final,3p,times,twocolumn]{elsarticle}
\biboptions{comma,sort&compress}
\usepackage{ecrc}
\usepackage{here}
\usepackage{graphicx}
\usepackage{epsfig}
\usepackage{epstopdf}
\usepackage{amsmath}

\def\nin{\noindent}
\def\beq{\begin{equation}}
\def\eeq{\end{equation}}
\def\bea{\begin{eqnarray}}
\def\eea{\end{eqnarray}}
\def\nnb{\nonumber}
\def\la{\langle}
\def\ra{\rangle}
\def\ga{\left(}
\def\dr{\right)}

\usepackage{graphicx}
\usepackage{here}
\def\beq{\begin{equation}}
\def\eeq{\end{equation}}
\def\bea{\begin{eqnarray}}
\def\eea{\end{eqnarray}}
\def\bq{\begin{quote}}
\def\eq{\end{quote}}
\def\ve{\vert}
\def\nnb{\nonumber}
\def\ga{\left(}
\def\dr{\right)}

\def\nnb{\nonumber}
\def\la{\langle}
\def\ra{\rangle}
\def\nin{\noindent}
\def\ba{\begin{array}}
\def\ea{\end{array}}

\def\als{\alpha_s}

\def\gg2{ \la\alpha_s G^2 \ra}
\def\gg3{g^3f_{abc}\la G^aG^bG^c \ra}
\def\ggg4{\la\als^2G^4\ra}

\def\beq{\begin{equation}}
\def\enq{\end{equation}}
\def\beqa{\begin{eqnarray}}
\def\enqa{\end{eqnarray}}
\def\nnb{\nonumber}

\def\MeV{\nobreak\,\mbox{MeV}}
\def\GeV{\nobreak\,\mbox{GeV}}
\def\keV{\nobreak\,\mbox{keV}}

\newcommand{\rag}{\rangle}
\newcommand{\lag}{\langle}

\def\ln{\mbox{Log}}
\def\gg{\lag g^{2}_{s} G^2 \rag}
\def\ggg{\lag g^{3}_{s}G^3\rag}

\volume{00}
\firstpage{1}
\journalname{Nuclear and Particle Physics Proceedings }
\runauth{D. Rabetiarivony}
\jnltitlelogo{Nuclear and Particle Physics Proceedings }

\begin{document}
\begin{frontmatter}

\title{$Z_{c,b}$ -like states from QCD Laplace sum rules at NLO \tnoteref{text1}}


\author[label2,label3]{S. Narison\fnref{fn1}}
\fntext[fn1]{ICTP-Trieste consultant for Madagascar}
\ead{snarison@yahoo.fr}
\address[label2]{Laboratoire Univers et Particules de Montpellier, CNRS-IN2P3, Case 070, Place Eug\`ene Bataillon, 34095 - Montpellier, France.}
\address[label3]{Institute of High Energy Physics of Madagascar (iHEPMAD), University of Ankatso, Antananarivo 101, Madagascar.}

\author[label3]{D. Rabetiarivony\fnref{fn2}}
\fntext[fn2]{Speaker}
\ead{rd.bidds@gmail.com} 


\tnotetext[text1]{Talk given at QCD21 International Conference (5--9 July 2021, Montpellier--FR)}

\pagestyle{myheadings}
\markright{ }

\begin{abstract}
\noindent
We review our results on $Z_{c}$-like states\,\cite{ANR2} which we complete with the ones on $Z_{b}$-like states by using relativistic QCD Laplace Sum Rules (LSR) within stability criteria and including Factorized Next-to-Leading Order (FNLO) Perturbative (PT) corrections
and  Lowest Order (LO) QCD condensates up to $\lag G^3 \rag$. We  emphasize the importance of PT radiative corrections for heavy quark sum rules in order to justify the use of the running heavy quark mass in the analysis. Our estimates are compiled in Tables \ref{tab:c-results} and \ref{tab:b-results}. From our results, the observed $Z_{cs}(3983)$ state are good candidate for being ${\cal T}_{cs}$ tetramole (superposition of nearly degenerated molecules and tetraquark states having the same quatum numbers $J^{PC}$ and with almost the same couplings to the currents). The $Z_{cs}$ bump around 4100 $\MeV$ can be interpreted as a combination of $D^{*}_{0}D_{s1}$ and $D^{*}_{s0}D_{1}$ molecules. The physical states $Z_{cs}(4000)$ and $Z_{cs}(4220)$ found by LHCb are too low to be considered as the first radial excitations of $Z_{cs}(3983)$. For the future $Z_{b}$, $Z_{bs}$ and $Z_{bss}$, we suggest to scan the region around $(10.3 \sim 10.9)$ $\GeV$ while the 1st radial excitations are about 2.4 GeV above the ground states.
\end{abstract} 
\scriptsize
\end{frontmatter}
\section{Introduction}
\nin In earlier papers \cite{AFNR,SNX1,SNX2,SU3,ANRR1,ANRR1a,QQQQ,ANRR2,ANR1}, the inverse Laplace Transform (LSR) \cite{BELL,BELLa,BNR,BERT,NEUF,SNR} of QCD spectral sum rules (QSSR)\footnote{For reviews, see \cite{SVZa,Za,SNB1,SNB2,SNB3,CK,YND,PAS,RRY,IOFF,DOSCH}} has been used to extract the masses and couplings of some heavy-light and fully heavy molecules and tetraquark states.

\nin In our recent work \cite{ANR2} reviewed here, we revisit our estimation for the $Z_c(3900)$ state found by BELLE\,\cite{BELLE1} and BESIII\,\cite{BES1} and present new results on $1^+$ $(\bar{c} s)(c \bar{u})$ molecules and $(c s)(\bar{c} \bar{u})$ tetraquark states for an attempt to interpret the recent observations of the $Z_{cs}$ physical states by BESIII\,\cite{BES2} and LHCb\,\cite{LHCb}. We have extended our predictions for their future $Z_{css}$ partner.\\
\nin In this talk, we pursue the analysis using LSR for the cases of future $Z_{b},~Z_{bs}$ and $Z_{bss}$.

\section{The Laplace sum rule}
\nin We shall work with the finite energy version of the QCD inverse Laplace sum rules and their ratios:
\bea
\mathcal{L}^{c}_{n}(\tau ,\mu)&=&\int^{t_c}_{t_0}dt\, t^n e^{-t\tau}\frac{1}{\pi}\mbox{Im}\Pi^{(1)}_{\mathcal{H}}(t,\mu),\nnb\\
\mathcal{R}^{c}_{n}(\tau)&=&\frac{\mathcal{L}^{c}_{n+1}}{\mathcal{L}^{c}_{n}},
\label{eq:LSR}
\eea
where $m_Q$ and $m_q$ are respectively the heavy and light quarks masses ($Q\equiv c,b$ and $q\equiv u,s$), $\tau$ is the LSR variable, $n=0,1$ is the degree of moments, $t_0$ is the hadronic threshold, $t_c$ is the "QCD continuum" which parametrizes, from the discontinuity of the Feynman diagrams, the spectral function $\mbox{Im}\Pi^{(1)}_{\mathcal{H}}(t,m^{2}_{Q},\mu^2)$ where $\Pi^{(1)}_{\mathcal{H}}(t,m^{2}_{Q},\mu^2)$ is the two-point scalar correlator defined as:
\bea
\hspace*{-0.6cm}
\Pi^{\mu\nu}_{\mathcal{H}}(q^2)\hspace*{-0.2cm}&=& \hspace*{-0.2cm} i \int \hspace*{-0.1cm} d^4 x\, e^{-i q x}\lag 0 \ve \mathcal{T} \mathcal{O}^{\mu}_{\mathcal{H}}(x)(\mathcal{O}^{\nu}_{\mathcal{H}}(x))^{\dag} \ve 0 \rag,\nnb \\
&\equiv &\hspace*{-0.2cm} -\left( g^{\mu\nu}-\frac{q^{\mu}q^{\nu}}{q^2}\right)\Pi^{(1)}_{\mathcal{H}}(q^2)+\frac{q^{\mu}q^{\nu}}{q^2}\Pi^{(0)}_{\mathcal{H}}(q^2),
\eea
where $\mathcal{O}^{\mu}_{\mathcal{H}}(x)$ are the interpolating currents desribing the molecules and tetraquark states.
\vspace*{-0.3cm}
\section{Interpolating currents}
\nin
We shall be concerned with the interpolating currents given in Table \ref{tab:current}.
\vspace*{-0.25cm}
\begin{table}[hbt]
\setlength{\tabcolsep}{0.55pc}
\caption{{\small Interpolating currents describing the $1^+$ molecules and tetraquark states ($Q\equiv c,b;~q\equiv u,s $}).
}
\begin{center}
{\small
\begin{tabular}{ll}
\hline
\hline
$1^+$ states& Currents ($\mathcal{O}^{\mu}_{\mathcal{H}}$) \\
\hline
{\bf Molecules}&\\
$D^{*}_{q} D,~B^{*}_{q} B $ & $(\overline{Q}\,\gamma_{\mu}\, q)(\overline{u}\, i\gamma_5\, Q)$\\
$D^{*} D_q,~B^{*} B_q $ & $(\overline{u}\,\gamma_{\mu}\, Q)(\overline{Q}\, i\gamma_5\, q)$\\
$D^{*}_{q0} D_{1},~B^{*}_{q0} B_1 $ & $(\overline{Q}\,q)(\overline{u}\, \gamma_{\mu}\gamma_5\, Q)$\\
$D^{*}_{0} D_{q1},~B^{*}_{0} B_{q1} $ & $(\overline{u}\,Q)(\overline{Q}\, \gamma_{\mu}\gamma_5\, q)$\\
$D^{*}_{s} D_s,~B^{*}_{s} B_s $ & $(\overline{Q}\,\gamma_{\mu}\, s)(\overline{s}\, i\gamma_5\, Q)$\\
$D^{*}_{s0} D_{s1},~B^{*}_{s0} B_{s1} $ & $(\overline{Q}\,s)(\overline{s}\, \gamma_{\mu}\gamma_5\, Q)$\\
{\bf Tetraquarks}&\\
$A_{cq},~A_{bq}$ & 
$\epsilon_{ijk}\,\epsilon_{mnk}\,\left[( q^{T}_{i}\, C\,\gamma_5\, Q_j ) (\bar{u}_m\, \gamma^\mu\,C\, \bar{Q}^{T}_{n})\right.$\\
& $ ~~~~~~~~~~~~~~ + \left. b\,( q^{T}_{i}\, C\, Q_j ) (\bar{u}_m\, \gamma^\mu\gamma_5\,C\, \bar{Q}^{T}_{n})\right]$ \\
$A_{css},~A_{bss}$ & 
$\epsilon_{ijk}\,\epsilon_{mnk}\,\left[( s^{T}_{i}\, C\,\gamma_5\, Q_j ) (\bar{s}_m\, \gamma^\mu\,C\, \bar{Q}^{T}_{n})\right.$\\
&$ ~~~~~~~~~~~~~~ + \left. b\,( s^{T}_{i}\, C\, Q_j ) (\bar{s}_m\, \gamma^\mu\gamma_5\,C\, \bar{Q}^{T}_{n})\right]$\\
\hline\hline
\end{tabular}
}
\end{center}
\label{tab:current}
\end{table}

\vspace*{-0.75cm}
\section{The Spectral function}
\nin
We shall use the Minimal Duality Ansatz (MDA) for parametrizing the molecule spectral function:
\beq
\frac{1}{\pi}\mbox{Im}\Pi_{{\cal H}}\hspace*{-0.1cm}\simeq \hspace*{-0.1cm} f^2_{{\cal H}}M^8_{{\cal H}}\delta(t-M_{{\cal H}}^2)+\Theta(t-t_c)\frac{1}{\pi}\mbox{Im}\Pi^{QCD}_{{\cal H}}(t),
\label{eq:mda}
\eeq
The hadron decay constant $f_{{\cal H}}$ (analogue to $f_{\pi}$) is defined as:
\bea
\lag 0 \ve \mathcal{O}^{\mu}_{\mathcal{H}} \ve \mathcal{H} \rag = f_{\mathcal{H}}\, M^{5}_{\mathcal{H}}\epsilon^{\mu}.
\label{eq:coupling}
\eea
Where $\epsilon^{\mu}$ is the axial-vector polarization.

\nin As the interpolating currents are constructed from bilinear (pseudo)scalar currents, thus the decay constants acquire anomalous dimension:
\bea
f_{\mathcal{H}}(\mu)=\hat{f}_{\mathcal{H}}(-\beta_1 \,a_s)^{2/\beta_1}(1-k_f\,a_s),
\eea
where $\hat{f}_{\mathcal{H}}$ is the renormalization group invariant coupling and $[-\beta_1=(1/2)(11-2n_f/3)]$ is the first coefficient of the QCD $\beta$-function for $n_f$ flavors. $a_s\equiv (\alpha_s/\pi)$ is the QCD coupling. $k_f=1.014(1.176)$ for $n_f=4(5)$ flavors.\\
\nin
Within a such parametrization, one obtains:
\bea
\mathcal{R}^{c}_{0}\equiv \mathcal{R}\simeq M^{2}_{\mathcal{H}},
\eea
where $M_{\mathcal{H}}$ is the lowest ground state mass.
\section{NLO PT corrections and stability criteria}
\label{sec:conv}
\nin
Assuming a factorization of the four-quark interpolating current,
we can write the corresponding spectral function as a convolution of the two ones associated to two quark bilinear currents.
In this way, we obtain \cite{PICH,NPIV}:

\bea
\hspace*{-0.7cm}
\frac{1}{ \pi}{\rm Im} \Pi_{\mathcal{H}}(t)\hspace*{-0.3cm}&=&\hspace*{-0.3cm}\theta \left(t-(\sqrt{t_{10}}+\sqrt{t_{20}})^2\right)  \ga \frac{k}{ 4\pi} \dr ^2 \hspace*{-0.15cm}  t^2 \hspace*{-0.15cm} \int_{t_{10}}^{(\sqrt{t}-\sqrt{t_{20}})^2} \hspace*{-0.2cm} dt_1 \nnb\\
&\times &\hspace*{-0.3cm} \int_{t_{20}}^{(\sqrt{t}-\sqrt{t_1})^2} \hspace{-0.2cm} dt_2 \,\left[ 2 \lambda^{3/2}\left(\frac{t_1}{t},\frac{t_2}{t}\right)\, \frac{1}{\pi}\mbox{Im}\psi^{S,P}(t_1)\right. \nnb\\
&& ~~~~~~~~~~~~~~~~~~~~~ \times  \left.\frac{1}{\pi}\mbox{Im}\Pi^{A,V}(t_2)\right],
\label{eq:conv}
\eea
where $\sqrt{t_{10}}$ and $\sqrt{t_{20}}$ are the quark/hadronic thresholds, $k$ is an appropriate normalization factor \cite{ANR2}, $m_Q$ is the on-shell heavy quark mass and with the phase space factor:
{\small
\beq
\lambda \left(\frac{t_1}{t},\frac{t_2}{t}\right) =\ga 1-\frac{\ga \sqrt{t_1}- \sqrt{t_2}\dr^2}{ t}\dr \ga 1-\frac{\ga \sqrt{t_1}+ \sqrt{t_2}\dr^2}{ t}\dr~.
\eeq
}
\nin
The NLO expressions of the spectral functions of the bilinear unequal masses (pseudo)scalar and (axial-)vector are known in the literature \cite{SNB1,SNB2,RRY,DJB,CS,GKPR}.\\
\nin The variables $\tau,\mu$ and $t_c$ are, in principle, free external parameters. We shall use stability criteria with respect to these free 3 parameters to extract the lowest ground state mass and coupling (more detailed discussions can be seen in \cite{AFNR,SNX1,SNX2,SU3,ANRR1,ANR1,ANR2,ADKT} and references therein).
\vspace*{-0.3cm}
\section{The On-shell and $\overline{\mbox{MS}}$-scheme}
\nin In our analysis, we replace the on-shell (pole) masses $m_Q$ appearing in the spectral functions with the running masses $\overline{m}_Q(\mu)$ using the relation, to order $\alpha_s$\,\cite{SNB1,SNB2}:
\bea
\hspace*{-0.5cm}
m_Q &=&\overline{m}_Q(\mu)\left[1+\frac{4}{3} a_s+ a_s\,\ln{\ga\frac{\mu}{ m_Q}\dr^2}+{\cal O}(a^{2}_{s})\right]
\label{eq:msb}
\eea
for $n_l$ light flavours where $\mu$ is the arbitrary subtraction scale.
\vspace*{-0.3cm}
\section{QCD input parameters}
\nin
The QCD parameters which shall appear in the following analysis will be the QCD coupling $\alpha_s$ \cite{SN18}, the charm and bottom quark masses $m_{c,b}$, the strange quark mass $m_s$ (we shall neglect the light quark masses $m_{u,d}$), the light quark condensate $\lag \overline{q}q\rag$ ($q\equiv u,d,s$), the gluon condensates $ \lag\alpha_sG^2\rag \equiv \la \alpha_s G^a_{\mu\nu}G_a^{\mu\nu}\ra$ and $ \la g^3G^3\ra \equiv \la g^3f_{abc}G^a_{\mu\nu}G^b_{\nu\rho}G^c_{\rho\mu}\ra$, the mixed condensate $\lag \overline{q}Gq\rag\equiv \lag \overline{q} g \sigma^{\mu\nu} (\lambda_a/2) G^{a}_{\mu\nu} q\rag= M^{2}_{0} \lag\overline{q} q\rag$ and the four-quark condensate $\rho \alpha_s \lag\overline{q} q \rag^2$, where $\rho\simeq 3-4$. Their values are given in Table \ref{tab:param}.
\begin{table}[hbt]
\setlength{\tabcolsep}{.5pc}
 \caption{{\small QCD input parameters from recent QSSR analysis based on stability criteria.}} 
{\small
 {\begin{tabular}{@{}llll@{}}
&\\
\hline
\hline
Parameters&Values& Ref.    \\
\hline
$\alpha_s(M_Z)$ & $0.1181(16)(3)$ & \cite{SN18,SN18a,SN181}\\
$\overline{m}_c(\overline{m}_c)$ [MeV] & $1266(16)$ & \cite{SN18,SN181,SN182,SN183,SN184}\\
$\overline{m}_b(\overline{m}_b)$ [MeV]& $4197(8)$ & \cite{SN182,SN183}\\
$\hat{\mu}_{q}$ [MeV] & $253(6)$ & \cite{SNB2,SN185}\\
$\hat{m}_s$ [MeV] & $114(6)$ & \cite{SNB2,SN185}\\
$\kappa\equiv \lag\overline{s}s\rag/\lag \overline{d}d\rag$ & $0.74(6)$ & \cite{SNB2,SN185,ANN}\\
$M^{2}_{0}$ [GeV$^2$] & $0.8(2)$ & \cite{SNB2,DOSCH,CDKS}\\
& &\cite{DJN,BIOFF,OPF,SN186}\\
$\la\alpha_s G^2\ra\times 10^2$ [GeV$^4$]& $6.35(35)$ & \cite{SN18,SN181}\\
$\la g^3  G^3\ra / \la\alpha_s G^2\ra$ [GeV$^2$]& $8.2(1.0)$ &\cite{SNB8a,SNB8b,SNB8c}\\
$\rho \alpha_s \lag\overline{q} q \rag^2\times 10^4$ [GeV$^6$] & $5.8(9)$ & \cite{DOSCH,LNT,SN187,DJN,BLR,CDKS}\\
\hline\hline
\end{tabular}}
}
\label{tab:param}
\end{table}
\vspace*{-0.75cm}
\section{Molecules and tetraquarks states}
\nin This section will be entirely devoted to the bottom channels. Detailed analyzes of their charmed analogue have already been carried out in \cite{ANR2} and the final results on $Z_{c}$ and $Z_{cs}$ states are quoted in Table \ref{tab:c-results}.

\nin As the analysis will be performed using the same techniques, we shall illustrate it in the case of $B^{*}_{s}B$. The final and conservative results are compiled in Table \ref{tab:b-results}.
\begin{table}[hbt]
\setlength{\tabcolsep}{.8pc}
\caption{{\small $Z_{c(s)}$ couplings and masses predictions from LSR at NLO. The errors from QCD input parameters are from Table \ref{tab:param}. We take $\ve \Delta \mu\ve=0.05$ GeV and $\ve \Delta \tau\ve =0.01$ GeV$^{-2}$. In the case of asymetric errors, we take the mean value.}}

{\small
 {\begin{tabular}{@{}lrr@{}}
&\\
\hline
\hline
Obsevables& Couplings [$\keV$]& Masses [$\MeV$]\\
\hline
\emph{Molecules} $(c \bar{d})(\bar{c} u)$&&\\
$D^{*} D$ & 140(15)&3912(61)\\
$D^{*}_{0} D_{1} $ &96(23)&4023(130)\\
\emph{Tetraquark} $(\bar{c}\bar{d})(c u)$&&\\
$A_{c}$ &173(17)&3889(58)\\
\emph{Molecules} $(\bar{c} s)(c \bar{u})$&&\\
$D^{*}_{s} D$ &130(15)&3986(51)\\
$D^{*} D_s$ & 133(16)&3979(56)\\
$D^{*}_{s0} D_{1} $ &86(23)&4064(133)\\
$D^{*}_{0} D_{s1}$ &89(22)&4070(133)\\
\emph{Tetraquark} $(c s)(\bar{c}\bar{u})$&&\\
$A_{cs}$ &148(17)&3950(56)\\
\hline\hline
\end{tabular}}
}
\label{tab:c-results}
\end{table}
\subsection{$B^{*}_{s}B$ states}
\begin{figure}[hbt] 
\begin{center}
{\includegraphics[width=3.8cm,height=2.6cm]{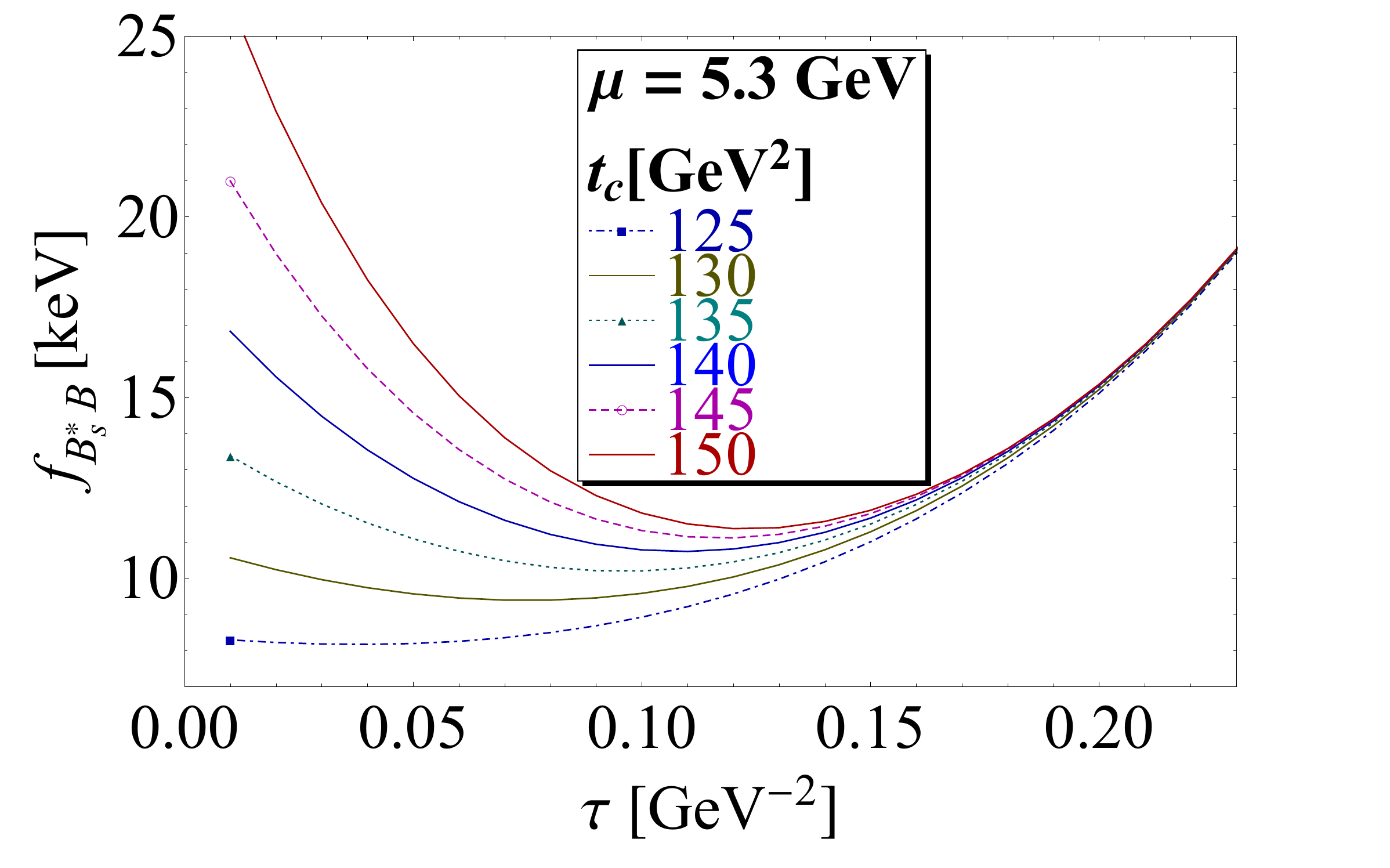}}
{\includegraphics[width=3.8cm,height=2.6cm]{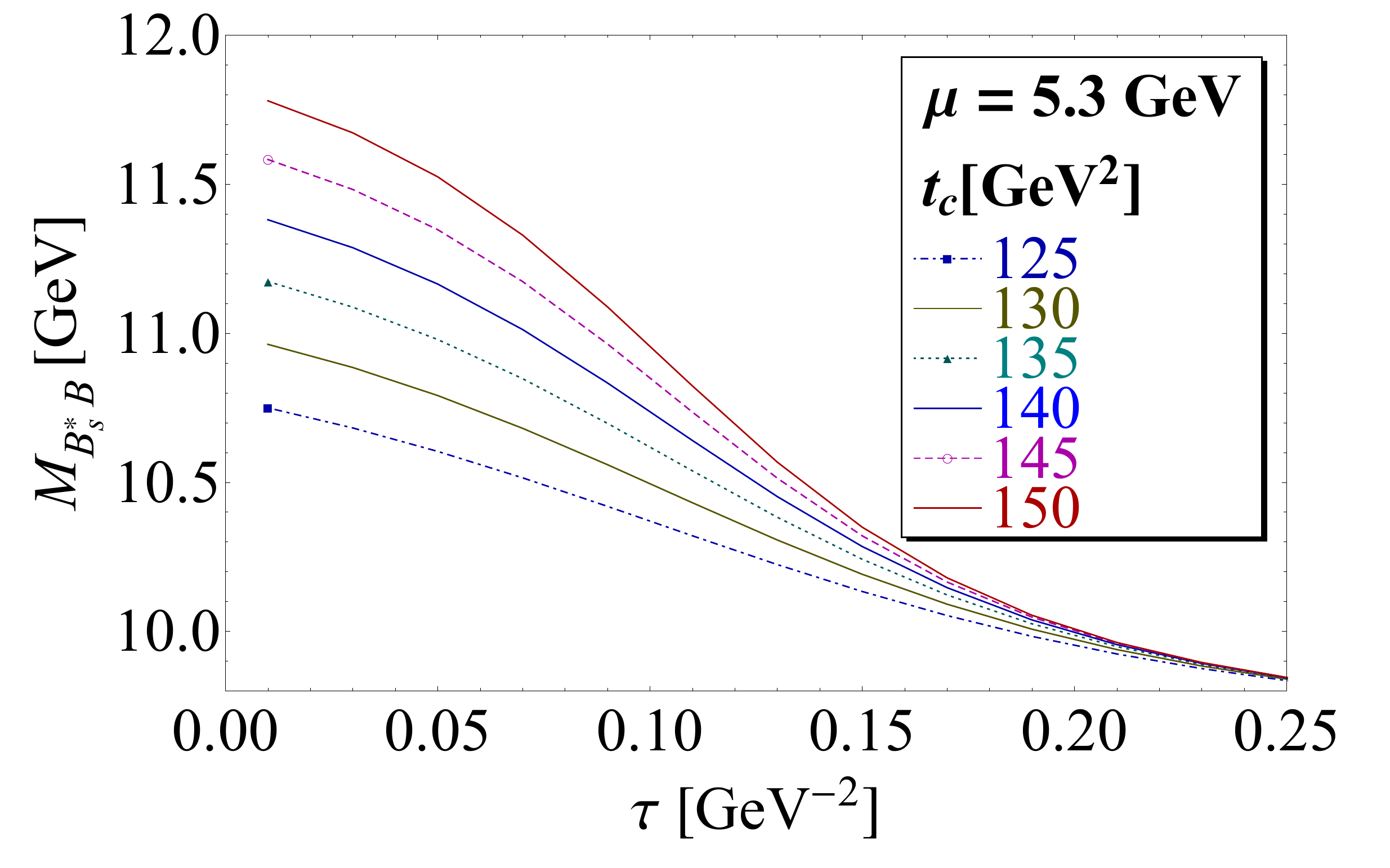}}
\caption{
{\small 
The coupling $f_{B^{*}_{s}B}$ and mass $M_{B^{*}_{s}B}$ at NLO as function of $\tau$ for different values of $t_c$, for $\mu=5.3$ GeV  and for the QCD parameters in Table\,\ref{tab:param}.}
}
\label{fig:bxsb}
\end{center}
\end{figure}

\nin
We study the behavior of the coupling and mass in term of the LSR variable $\tau$ for different values of $t_c$ at NLO as shown in Fig.\,\ref{fig:bxsb}.
\nin
We consider as final results the mean of the values corresponding to the beginning of $\tau$ stability for $[t_c\,(\GeV),\,\tau\,(\GeV^{-2})]\simeq [130,\,0.08]$ and the one where the $t_c$ stability is reached for $[t_c\,(\GeV),\,\tau\,(\GeV^{-2})]\simeq [145,\,0.12]$. For the other channels, the set of $(t_c,\tau)$ values where the optimal results have been extracted are summarized in Table \ref{tab:b-param}.
\vspace*{-0.25cm}
\subsection{$\mu$-stability}
\begin{figure}[hbt] 
\begin{center}
{\includegraphics[width=3.8cm,height=2.6cm]{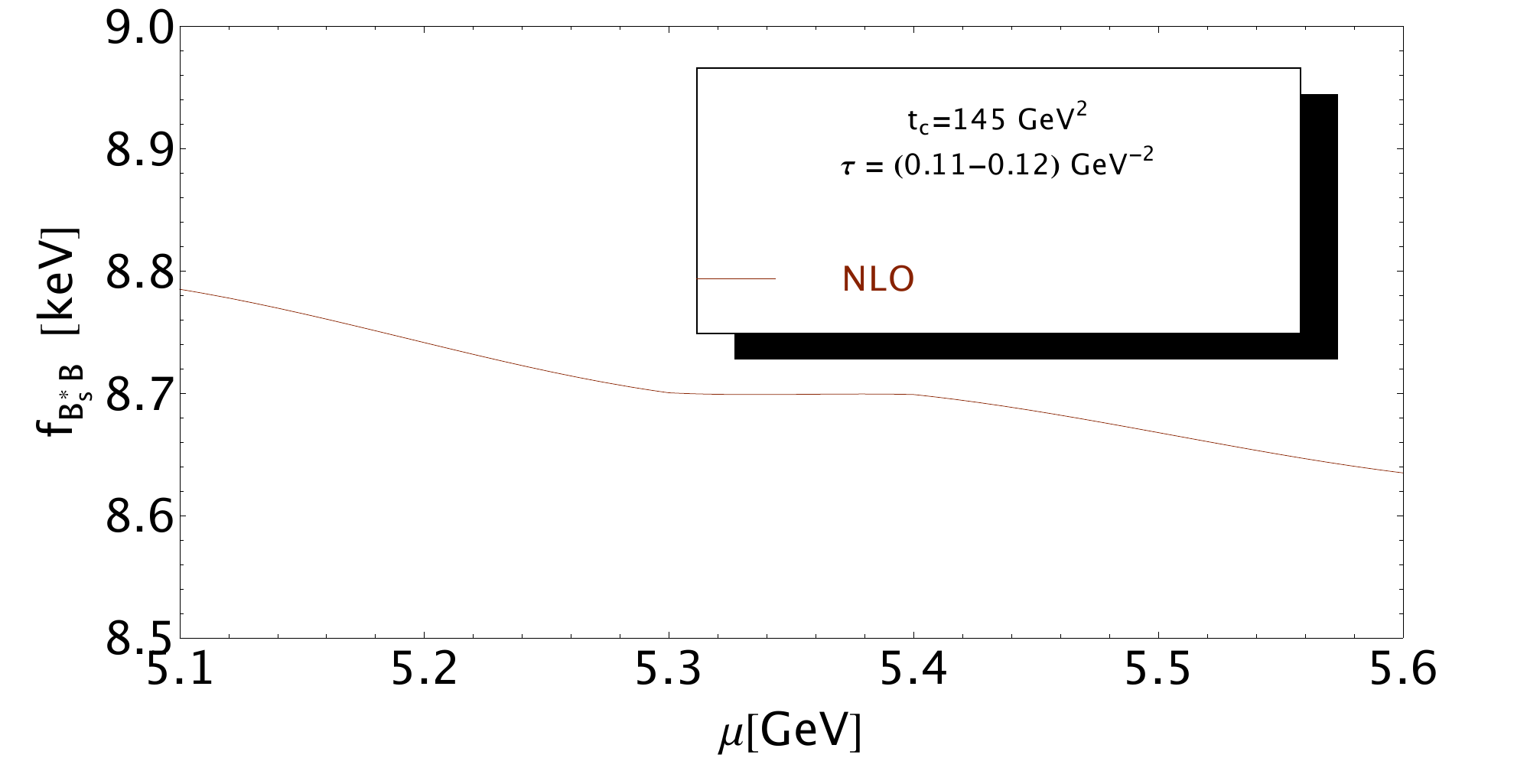}}
{\includegraphics[width=3.8cm,height=2.6cm]{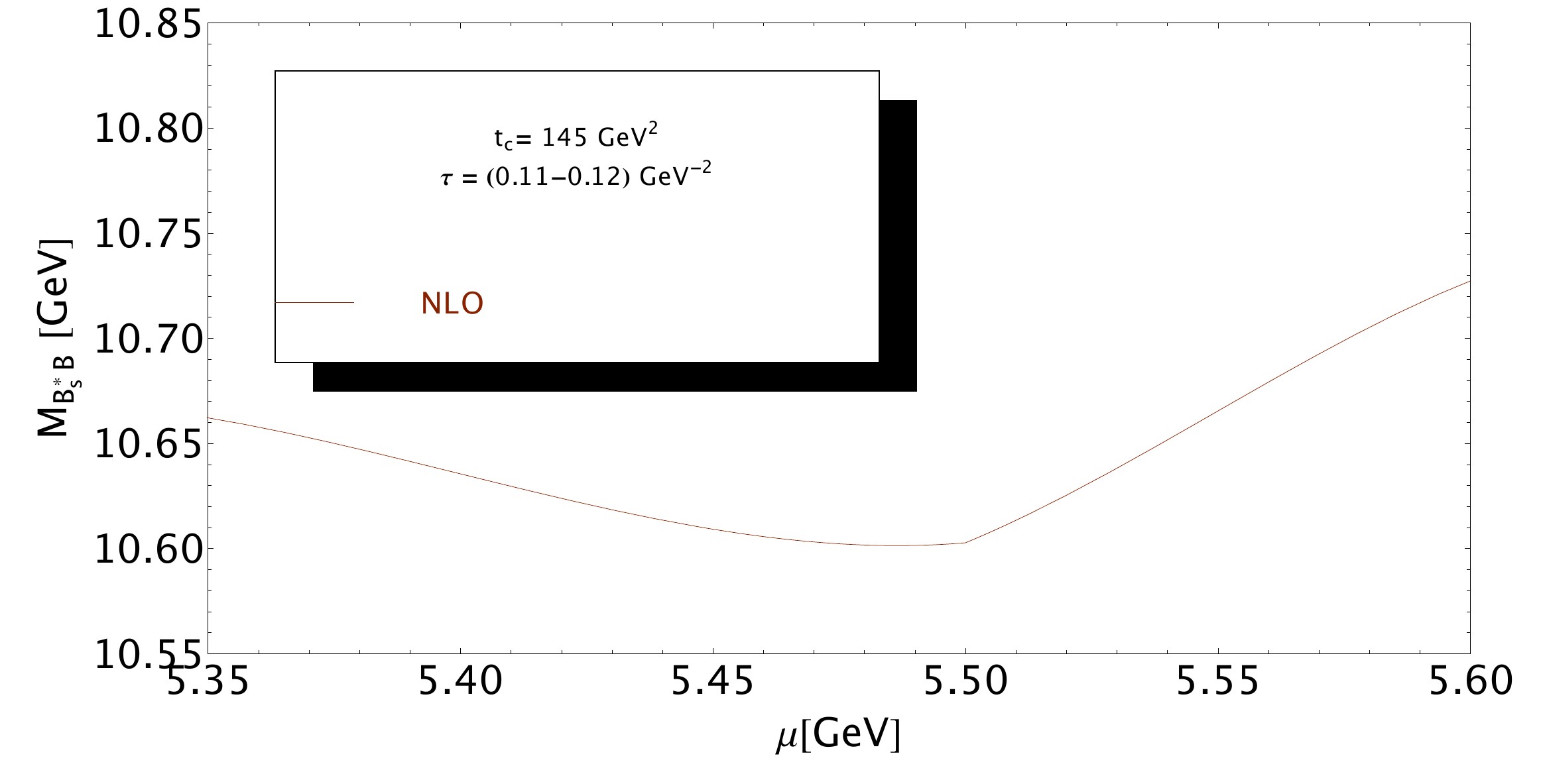}}
\caption{
{\small 
$\mu$-behavior of $f_{B^{*}_{s}B}$ and $M_{B^{*}_{s}B}$ for $t_c=145$ GeV$^2$ at NLO.}
}
\label{fig:bxsb-mu}
\end{center}
\end{figure}

\nin
Using the fact that the final results must be independent of the arbitrary parameter $\mu$, we consider as optimal result the one at the inflexion point for $\mu\simeq 5.3$ GeV (Fig.\,{\ref{fig:bxsb-mu}}).
\vspace*{-0.5cm}
\subsection{The Factorization assumption}
\label{sec:fnf}
\begin{figure}[hbt] 
\begin{center}
{\includegraphics[width=3.8cm,height=2.6cm]{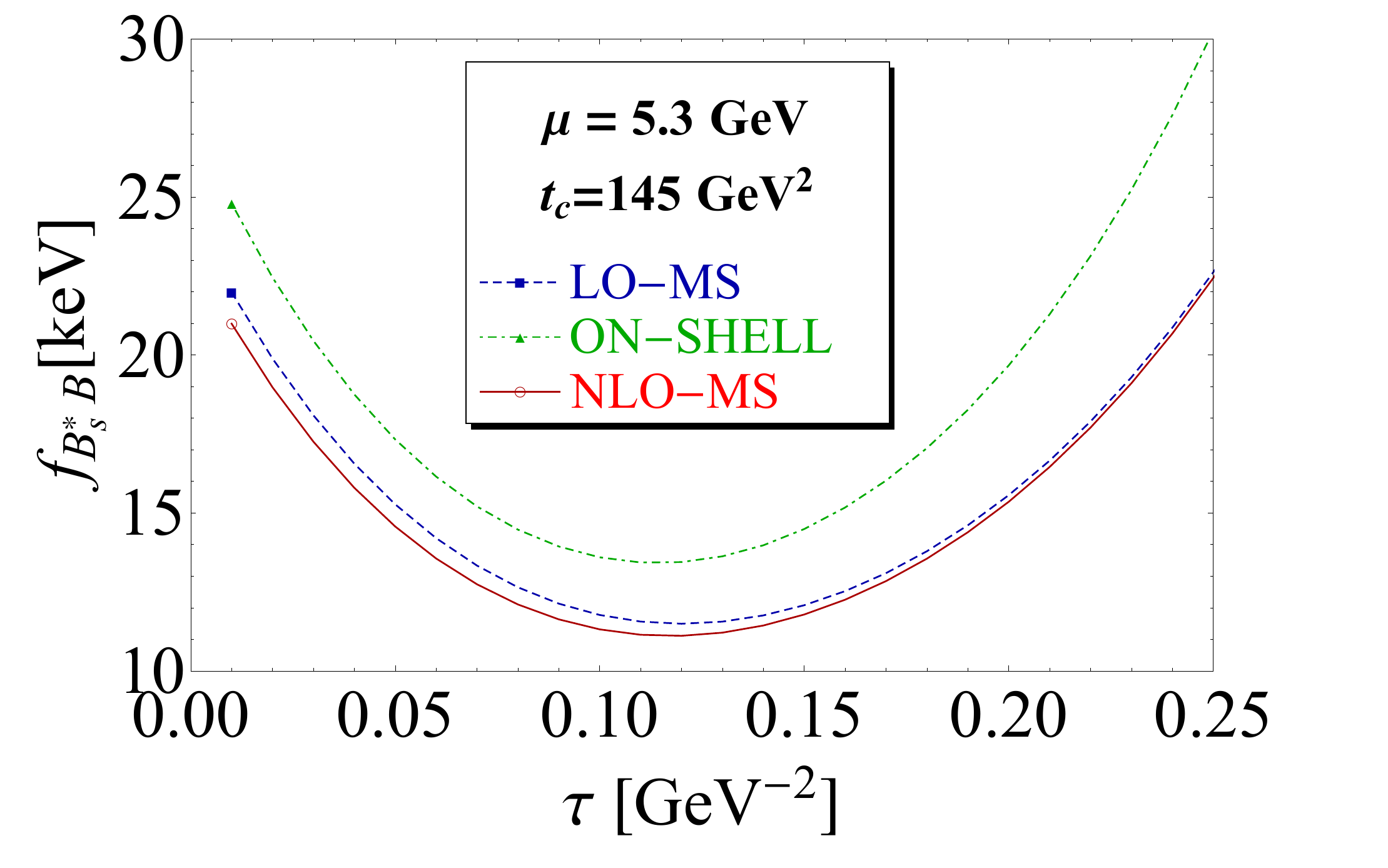}}
{\includegraphics[width=3.8cm,height=2.6cm]{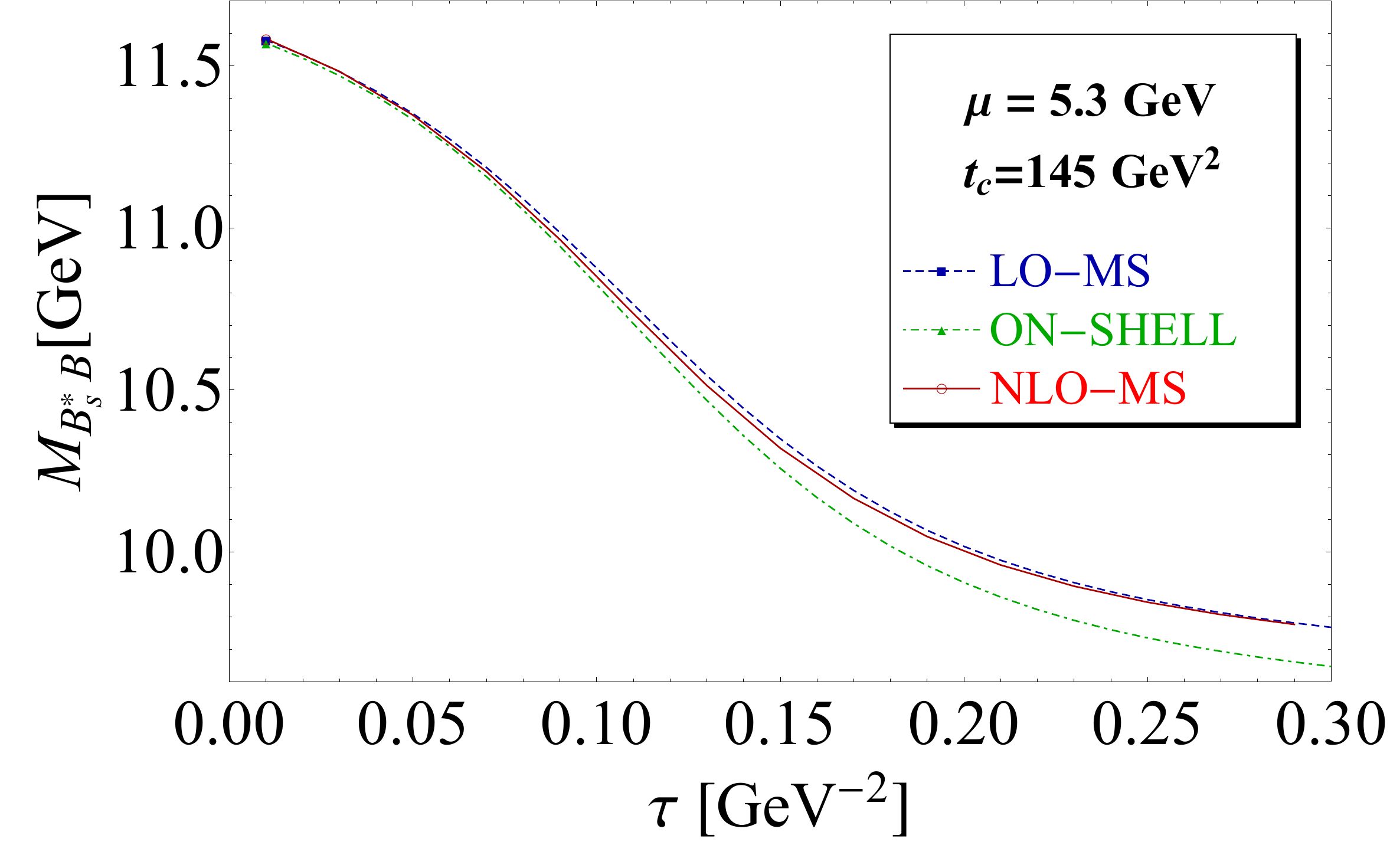}}
\caption{
{\small 
The coupling $f_{B^{*}_{s}B}$ and mass $M_{B^{*}_{s}B}$ at LO and NLO as function of $\tau$ for $t_c=145~\GeV^2$ and $\mu=5.3$ GeV for different definitions of the bottom quark mass. We use $m_b(on-shell)=4.66~\GeV$ and the running mass given in Table \ref{tab:param}.}
}
\label{fig:lo-nlo}
\end{center}
\end{figure}

\nin We have shown explicitly in \cite{ANR2} that the contributions of non-factorized part to the four-quark correlator which appear at LO ($\alpha^{0}_{s}$) of perturbative series but not at $\alpha^{2}_{s}$ (as claimed by \cite{LMS}) are numerically negligible. This feature occurs also in the cases of the bottom channels, as we can see in Fig.\,\ref{fig:lo-nlo} which justifies our approximation by considering the four-quark spectral function as a convolution of two ones built from two quark bilinear currents. One can also notice from Fig.\,\ref{fig:lo-nlo} that at LO, the two definitions of the bottom quark mass lead to different predictions.  However, using the $\overline{MS}$ running mass, the LO and NLO give almost the same results indicating that in this scheme the radiative corrections are small which justify the good numerical results obtained at LO in the current literature. These features have already been tested in our previous works \cite{SNX1,SNX2,SU3,ANRR1,ANRR1a,QQQQ}.
\nin
\vspace*{-0.75cm}
\subsection{The Tetramole states}
\label{sec:tetra}
\nin We expect that the "physical states" is a combination of nearly degenerated molecules and tetraquark states having the same quantum numbers $J^{PC}$ with almost the same couplings strengths to the currents which we shall call \emph{Tetramole} (${\cal T}$). Taking the quadratic means of the masses and couplings of these states, we obtain:\\
-- $B^*B\oplus A_{bd}$\\
\beq
M_{{\cal T}_{b}}=10580(148)\MeV , ~~~~ f_{{\cal T}_{b}}=10(2)\keV
\eeq
-- $B^{*}_{s}B \oplus B^*B_s \oplus A_{bs}$\\
\beq
M_{{\cal T}_{bs}}=10644(132)\MeV , ~~~~ f_{{\cal T}_{bs}}=10(2)\keV
\eeq
-- $B^{*}_{s}B_{s} \oplus A_{bss}$\\
\beq
M_{{\cal T}_{bss}}=10703(195)\MeV , ~~~~ f_{{\cal T}_{bss}}=9(3)\keV.
\eeq
The value of $M_{{\cal T}_{b}}$ agrees with the one of\,\cite{AZIZI} obtained to LO  and just below the $t_c$-stability region (130-150 GeV$^2$) where the $\tau$-stability is not yet reached. This agreement is due to the fact that in the $\overline{MS}$ scheme the radiative corrections are small as shown in Fig.\,\ref{fig:lo-nlo}.

\subsection{Radial excitations}
\label{sec:radial}
\begin{figure}[hbt] 
\begin{center}
{\includegraphics[width=3.8cm,height=2.6cm]{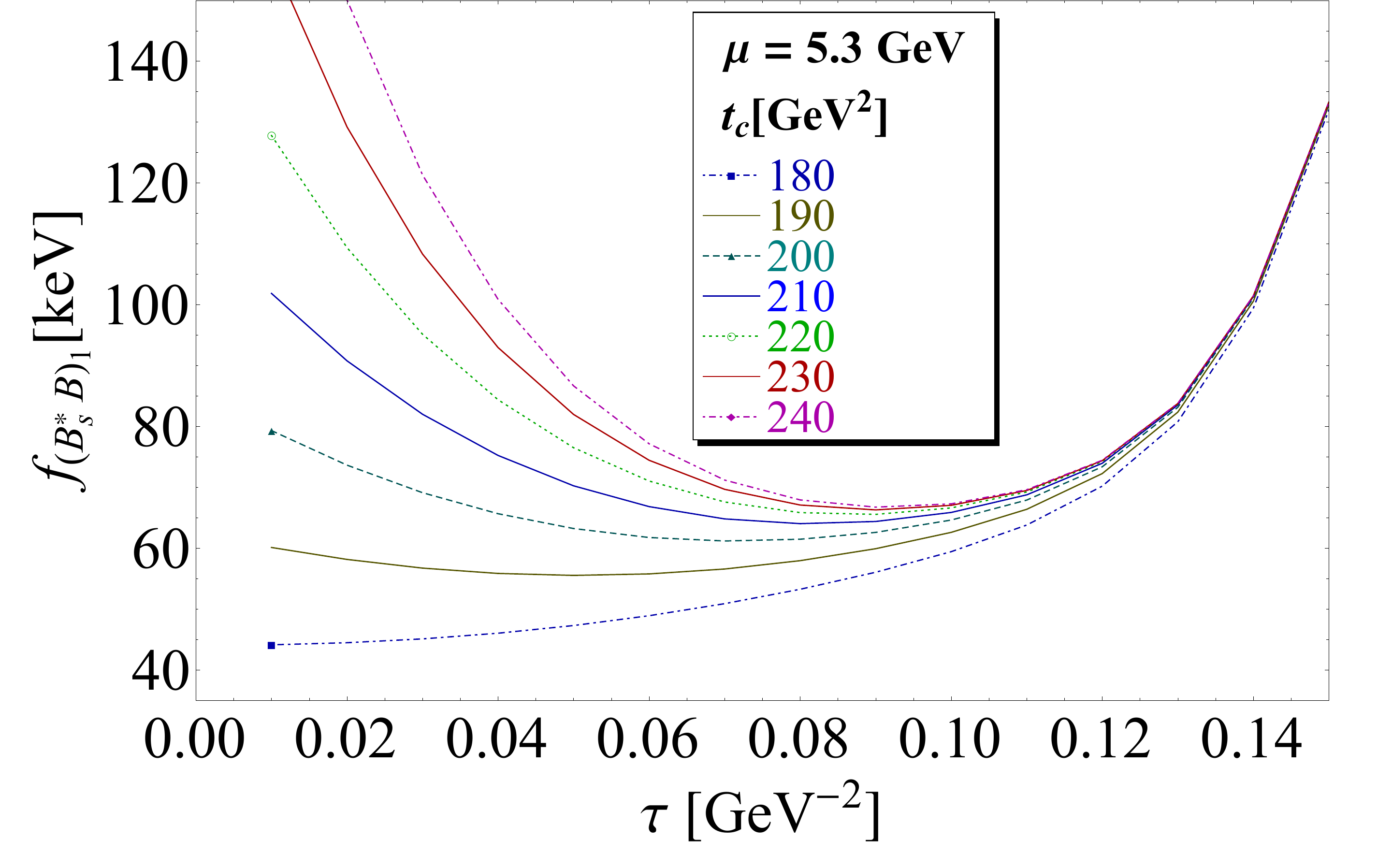}}
{\includegraphics[width=3.8cm,height=2.6cm]{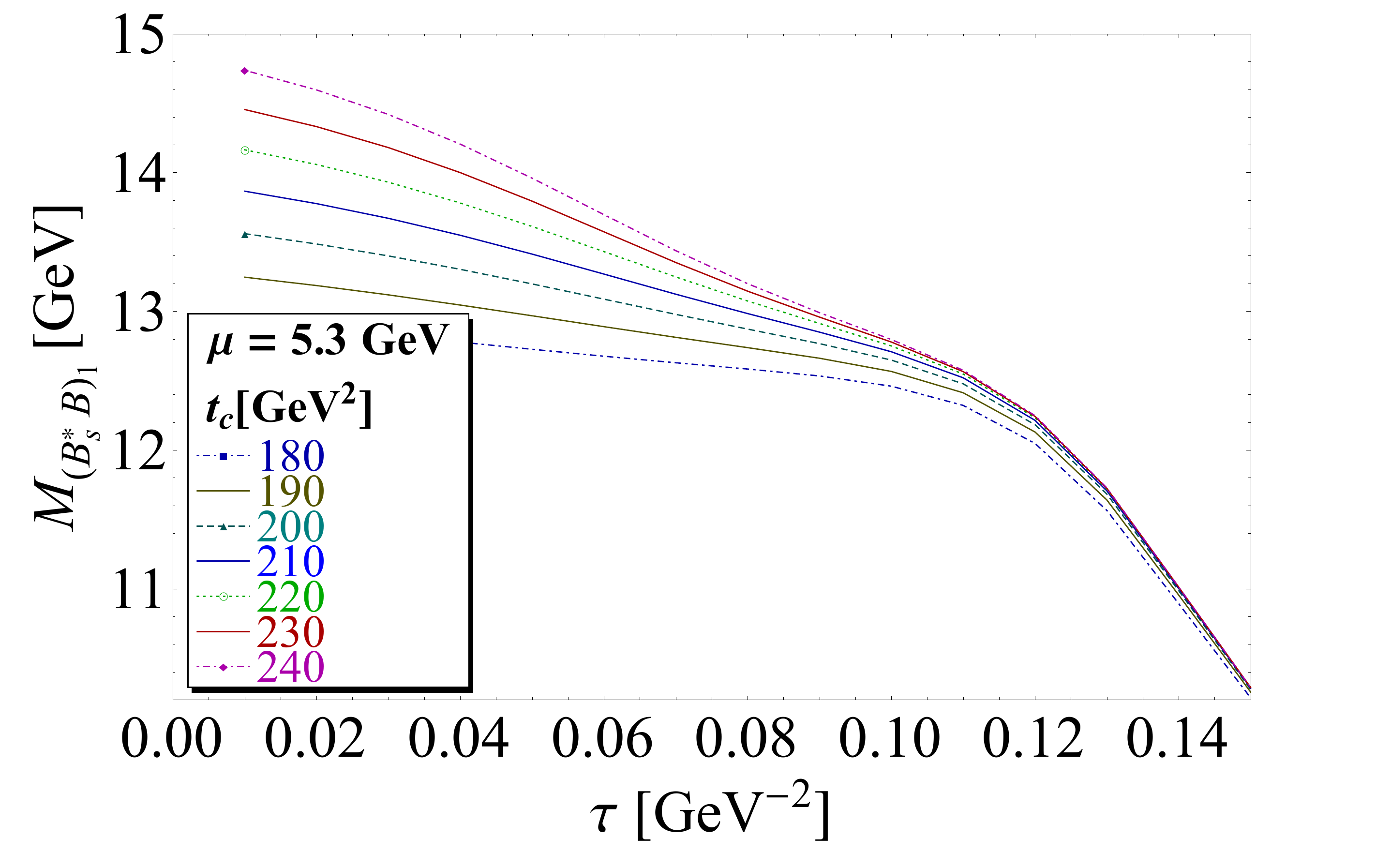}}
\caption{
{\small 
The coupling $f_{(B^{*}_{s}B)_1}$ and mass $M_{(B^{*}_{s}B)_1}$ at NLO as function of $\tau$ for different values of $t_c$, for $\mu=5.3$ GeV  and for the QCD parameters in Table\,\ref{tab:param}.}
}
\label{fig:bxsb-rad}
\end{center}
\end{figure}

\nin We estimate the couplings and masses of first radial excitations of $B^{*}B$, $B^{*}_{s}B$ and $B^{*}_{s}B_s$ using a "Two resonances"+$\theta(t-t_c)$ "QCD continuum" parametrization of the spectral function. We shall also work with the moments ${\cal L}^{c}_{1}$ and the ratio of moments ${\cal R}^{c}_{1}$. The behavior of the different curves are very similar to the case of $(B^{*}_{s}B)_1$ shown in Fig. \ref{fig:bxsb-rad} where the stability region is delimited by $t_c=190$ and $230\,\GeV^2$ to which corresponds respectively the $\tau$ stability of ($0.05\sim 0.06$)$\,\GeV^{-2}$ and ($0.09\sim 0.10$)$\,\GeV^{-2}$. The set of LSR parameters used to get the final results compiled in Table \ref{tab:res-rad} are quoted in Table \ref{tab:param-rad}. One can notice from Table \ref{tab:res-rad} that the couplings of the first radial excitations are relatively large compared to the ones of the lowest ground sates. We have shown also that the mass-splittings between the first rdial excitation and the lowest ground state are
\beq
M_{(G)_1}-M_{G}\approx 2.4\,\GeV.
\eeq
\vspace*{-1cm}
\vspace*{-1cm}
\section{Conclusions}
\vspace*{-0.25cm}
\nin
-- Motivated by the first observations of hidden-charm tetraquark with strangeness by BESIII and LHCb, we have presented new predictions of $(\bar{c} s)(c \bar{u})$ molecules and $(c s)(\bar{c} \bar{u})$ tetraquark states using LSR within stability criteria where we have included to the LO perturbative term, the NLO corrections and contributions of quark and gluon condensates up to dimension-six. We have completed the analysis with new predictions of the future beauty $Z_b,~Z_{bs}$ and $Z_{bss}$ states.\\
-- We have introduced the tetramole states as a superposition of nearly degenerated molecules and tetraquark states having the same quantum numbers $J^{PC}$ with almost the same couplings to the currents. The mass splittings of the tetramoles ${\cal T}_{Q},~{\cal T}_{Qs}, {\cal T}_{Qss}$ ($Q\equiv c,b$) due to $SU(3)$ breakings are successively about ($73\sim 91$) MeV (resp. $\sim 60$ MeV) for the charm (resp. bottom) channel.\\
-- The $Z_{cs}(3983)$ can be identified with the tetramole ${\cal T}_{cs}(3973)$ and the $Z_{cs}(4100)$ bump as a combination of $D^{*}_{s0}D_{1}$ and $D^{*}_{0}D_{s1}$ molecules.
\\
-- Our predictions suggest the presence of $Z_b,~Z_{bs}$ and $Z_{bss}$ in the range ($10.3\sim 10.9$) GeV which can be tested in future experiments while the 1st radial excitations are expected to be about 2.4 GeV above the ground states. 

\begin{table*}[H]
\setlength{\tabcolsep}{0.46pc}
 \caption{{\small $Z_{b(s)}$-like couplings and masses predictions from LSR at NLO. The errors from QCD input parameters are from Table \ref{tab:param}. We take $\ve \Delta \mu\ve=0.10$ GeV and $\ve \Delta \tau\ve =0.01$ GeV$^{-2}$. In the case of asymetric errors, we take the mean value.}}
{\scriptsize{
\begin{tabular*}{\textwidth}{@{}ll ll  ll  ll ll ll ll ll r@{\extracolsep{\fill}}l}
\hline
\hline
                \bf Observables &\multicolumn{1}{c}{$\Delta t_c$}
					&\multicolumn{1}{c}{$\Delta \tau$}
					&\multicolumn{1}{c}{$\Delta \mu$}
					&\multicolumn{1}{c}{$\Delta \alpha_s$}
					&\multicolumn{1}{c}{$\Delta PT$}
					&\multicolumn{1}{c}{$\Delta m_s$}
					&\multicolumn{1}{c}{$\Delta m_b$}
					&\multicolumn{1}{c}{$\Delta \overline{\psi}\psi$}
					&\multicolumn{1}{c}{$\Delta \kappa$}					
					&\multicolumn{1}{c}{$\Delta G^2$}
					&\multicolumn{1}{c}{$\Delta M^{2}_{0}$}
					&\multicolumn{1}{c}{$\Delta \overline{\psi}\psi^2$}
					&\multicolumn{1}{c}{$\Delta G^3$}
					&\multicolumn{1}{c}{$\Delta OPE$}
					&\multicolumn{1}{c}{$\Delta M_G$}
					&\multicolumn{1}{r}{Values}\\
\hline
{\bf Coupling} $f_{_G}$[$\keV$]&&&&&&&&&&&&&&\\
\emph{Molecules} $(b\bar{d})(\bar{b}u)$&&&&&&&&&&&&&&\\
$B^{*} B $ &0.80&0.05&0.02&0.29&0.47&$\cdots$&0.13&0.23&$\cdots$&0.0&0.14&0.49&0.0&1.18&0.70 & 9(2) \\
$B^{*}_{0} B_{1}$ &0.89&0.12&0.07&0.19&0.47&$\cdots$&0.14&0.33&$\cdots$&0.0&0.18&0.65&0.0&2.20&1.65&8(3)\\
\emph{Tetraquark} $(\bar{b}\bar{d})(b u)$&&&&&&&&&&&&&&&&\\
$A_{bd}$ &0.83&0.12&0.02&0.33&0.52&$\cdots$&0.14&0.25&$\cdots$&0.0&0.17&0.52&0.0&0.52&0.62&11(2)\\
\emph{Molecules} $(\bar{b} s)(b \bar{u})$&&&&&&&&&&&&&&&&\\
$B^{*}_{s} B$ &0.86&0.62&0.08&0.33&0.02&0.01&0.18&0.16&0.28&0.0&0.12&0.39&0.0&1.29&0.84&10(2) \\
$B^{*} B_s$ &0.85&0.87&0.07&0.34&0.01&0.0&0.19&0.16&0.27&0.0&0.13&0.38&0.0&1.34&0.95&10(2) \\
$B^{*}_{s0} B_{1} $ & 0.79&0.13&0.08&0.22&1.05&0.01&0.16&0.26&0.17&0.0&0.18&0.58&0.0&2.12&2.35&9(4)\\
$B^{*}_{0} B_{s1}$ &0.80&0.12&0.08&0.22&1.10&0.01&0.16&0.26&0.17&0.0&0.17&0.58&0.0&2.12&2.37&9(4) \\
\emph{Tetraquark} $(b s)(\bar{b}\bar{u})$&&&&&&&&&&&&&&&&\\
$A_{bs}$ &0.93&0.05&0.09&0.38&0.05&0.01&0.21&0.18&0.31&0.0&0.16&0.45&0.0&1.56&1.12&11(2) \\
\emph{Molecules} $(\bar{b} s)(b \bar{s})$&&&&&&&&&&&&&&&&\\
$B^{*}_{s} B_s$ &0.43&0.06&0.05&0.31&1.00&0.0&0.14&0.22&0.64&0.0&0.15&0.37&0.0&1.42&1.30&8(2) \\
$B^{*}_{s0} B_{s1}$ & 0.81&0.11&0.08&0.20&0.46&0.05&0.14&0.27&0.23&0.0&0.18&0.53&0.0&1.89&2.14&8(3)\\
\emph{Tetraquark} $(b s)(\bar{b}\bar{s})$&&&&&&&&&&&&&&&&\\
$A_{bss}$ &0.76&0.07&0.20&0.42&1.78&0.11&0.13&0.39&0.90&0.0&0.22&0.43&0.0&1.49&1.03&9(3)\\
{\bf Mass} $M_{_G}$[$\MeV$]&&&&&&&&&&&&&&&&\\
\emph{Molecules} $(b\bar{d})(\bar{b}u)$&&&&&&&&&&&&&&&&\\
$B^{*} B $ & 15.3&88.0&17.0&4.50&0.30&$\cdots$&6.90&15.4&$\cdots$&0.03&0.23&16.4&0.05&139&$\cdots$&10582(169)\\
$B^{*}_{0} B_{1}$ &33.6&255&5.60&12.8&0.22&$\cdots$&5.00&57.0&$\cdots$&0.03&2.10&70.0&0.05&149&$\cdots$&10626(311) \\
\emph{Tetraquark} $(\bar{b}\bar{d})(b u)$&&&&&&&&&&&&&&&&\\
$A_{bd}$ &14.2&89.0&17.5&5.00&0.31&$\cdots$&7.30&14.3&$\cdots$&0.05&1.00&56.7&0.0&56.0&$\cdots$&10578(123)\\

\emph{Molecules} $(\bar{b} s)(b \bar{u})$&&&&&&&&&&&&&&&&\\
$B^{*}_{s} B$ &29.0&84.6&30.6&14.7&0.0&14.9&14.7&14.8&14.1&0.05&14.9&16.5&0.05&68.5&$\cdots$&10652(123) \\
$B^{*} B_s$ &10.0&86.0&6.70&10.0&0.01&2.75&5.40&10.7&14.3&0.20&5.50&25.0&0.20&96.3&$\cdots$&10651(134)\\
$B^{*}_{s0} B_{1} $ &31.8&239&5.20&14.3&0.14&1.15&1.55&32.5&53.4&0.0&8.10&73.5&0.0&247&$\cdots$&10713(359) \\
$B^{*}_{0} B_{s1}$ &33.7&238&4.88&14.3&0.15&0.83&1.50&31.0&54.1&0.0&8.80&74.1&0.0&244&$\cdots$&10725(356) \\
\emph{Tetraquark} $(b s)(\bar{b}\bar{u})$&&&&&&&&&&&&&&&&\\
$A_{bs}$ &7.95&86.5&7.45&9.65&0.09&1.55&5.40&11.9&13.9&0.05&5.05&39.9&0.0&97.1&$\cdots$&10630(138)\\
\emph{Molecules} $(\bar{b} s)(b \bar{s})$&&&&&&&&&&&&&&&&\\
$B^{*}_{s} B_s$ &24.5&130&27.5&3.40&1.00&25.4&25.1&26.6&11.2&0.0&25.8&21.8&0.0&139&$\cdots$&10694(202)\\
$B^{*}_{s0} B_{s1}$ &13.0&253&3.50&12.1&1.04&2.33&3.48&35.8&83.0&0.03&12.4&80.8&0.03&248&$\cdots$&10773(375)\\
\emph{Tetraquark} $(b s)(\bar{b}\bar{s})$&&&&&&&&&&&&&&&&\\
$A_{bss}$ &4.85&148&22.2&32.1&0.68&13.6&3.33&40.0&17.7&0.18&2.00&23.6&0.03&95.0&$\cdots$&10711(188)\\
\hline
\hline
\end{tabular*}
}}
\label{tab:b-results}
\end{table*}
\begin{table*}[H]
\setlength{\tabcolsep}{0.34pc}
 \caption{{\small Radial excitations of some $Z_{b(s)}$ states. The errors from QCD input parameters are from Table \ref{tab:param}. We take $\ve \Delta \mu\ve=0.10$ GeV and $\ve \Delta \tau\ve =0.01$ GeV$^{-2}$.}}
{\scriptsize{
\begin{tabular*}{\textwidth}{@{}ll ll  ll  ll ll ll ll ll ll r@{\extracolsep{\fill}}l}
\hline
\hline
                \bf Observables &\multicolumn{1}{c}{$\Delta t_c$}
					&\multicolumn{1}{c}{$\Delta \tau$}
					&\multicolumn{1}{c}{$\Delta \mu$}
					&\multicolumn{1}{c}{$\Delta \alpha_s$}
					&\multicolumn{1}{c}{$\Delta PT$}
					&\multicolumn{1}{c}{$\Delta m_s$}
					&\multicolumn{1}{c}{$\Delta m_b$}
					&\multicolumn{1}{c}{$\Delta \overline{\psi}\psi$}
					&\multicolumn{1}{c}{$\Delta \kappa$}					
					&\multicolumn{1}{c}{$\Delta G^2$}
					&\multicolumn{1}{c}{$\Delta M^{2}_{0}$}
					&\multicolumn{1}{c}{$\Delta \overline{\psi}\psi^2$}
					&\multicolumn{1}{c}{$\Delta G^3$}
					&\multicolumn{1}{c}{$\Delta OPE$}
					&\multicolumn{1}{c}{$\Delta M_G$}
					&\multicolumn{1}{c}{$\Delta f_G$}
					&\multicolumn{1}{c}{$\Delta M_{(G)_1}$}
					&\multicolumn{1}{r}{Values}\\
\hline
{\bf Coupling} $f_{_G}$[$\keV$]&&&&&&&&&&&&&&&\\
$(B^{*} B)_1 $ &4.61&0.55&0.41&1.96&3.02&$\cdots$&0.87&1.50&$\cdots$&0.0&0.56&2.14&0.0&3.93&1.89&4.03&5.31&56(10)& \\
$(B^{*}_{s} B)_1$ &5.38&0.53&0.24&2.13&0.38&0.12&1.24&1.11&1.33&0.0&0.49&1.68&0.0&3.30&1.81&3.96&5.98&61(10) \\
$(B^{*}_{s} B_s)_1$ &2.92&0.35&0.99&4.93&3.24&0.24&3.39&2.87&3.25&0.0&1.53&6.09&0.0&3.75&2.80&4.73&10.6&42(17) \\
{\bf Mass} $M_{_G}$[$\MeV$]&&&&&&&&&&&&&&&&&&\\
$(B^{*} B)_1 $ & 17.1&117&26.4&64.2&0.05&$\cdots$&62.8&19.3&$\cdots$&0.0&59.0&41.3&0.10&103&26.8&56.3&$\cdots$&12942(207)\\
$(B^{*}_{s} B)_1$ &4.85&132&18.4&35.4&0.14&1.03&47.6&34.1&52.7&0.13&27.1&16.2&0.20&67.4&30.7&30.4&$\cdots$&12963(181) \\
$(B^{*}_{s} B_s)_1$ &23.6&150&4.28&6.45&0.01&4.53&58.2&85.3&97.0&0.03&63.7&17.0&0.20&45.9&34.1&45.4&$\cdots$ &13020(230)\\
\hline
\hline
\end{tabular*}
}}
\label{tab:res-rad}
\end{table*}
\begin{table*}[H]
\setlength{\tabcolsep}{0.6pc}
 \caption{{\small Values of the set of LSR parameters $(t_c,\tau)$ at the optimization region for the PT series up to NLO and for the OPE truncated at the dimension-six condensates and for $\mu=5.3$ $\GeV$.}}
{\scriptsize{
\begin{tabular*}{\textwidth}{@{}ll ll  ll  ll ll ll ll ll r@{\extracolsep{\fill}}l}
\hline
\hline
                \bf States &\multicolumn{1}{c}{$B^{*}B$}
					&\multicolumn{1}{c}{$B^{*}_{s}B$}
					&\multicolumn{1}{c}{$B^{*}B_{s}$}
					&\multicolumn{1}{c}{$B^{*}_{s}B_{s}$}
					&\multicolumn{1}{c}{$B^{*}_{0}B_1$}
					&\multicolumn{1}{c}{$B^{*}_{s0}B_1$}
					&\multicolumn{1}{c}{$B^{*}_{0}B_{s1}$}
					&\multicolumn{1}{c}{$B^{*}_{s0}B_{s1}$}
					&\multicolumn{1}{c}{$A_{bd}$}
					&\multicolumn{1}{c}{$A_{bs}$}
					&\multicolumn{1}{c}{$A_{bss}$}					
					
                  \\
\hline
\bf Parameters&&&&&&&&&&&\\
$t_{c}$ $[\GeV]^2$&130-150&130-145&130-145&145-165&145-165&145-165&145-165&145-165&130-150&130-145&150-170\\
$\tau$ $[\GeV]^{-2}\times 10^2$&$~~5~~;~10$&$~~7~~;~12$&$~~8~~;~12$&$~~9~~;~12$&$~~8~~;~11$&$~~9~~;~12$&$~~9~~;~12$&$~~9~~;~12$&$~~6~~;~10$&$~~8~~;~12$&$~~7~~;~11$\\
\hline
\hline
\end{tabular*}
}}
\label{tab:b-param}
\end{table*}
\begin{table*}[H]
\setlength{\tabcolsep}{4pc}
 \caption{{\small Values of the set of LSR parameters $(t_c,\tau)$ at the optimization region for the PT series up to NLO and for the OPE truncated at the dimension-six condensates and for $\mu=5.3$ $\GeV$.}}
{\scriptsize{
\begin{tabular*}{\textwidth}{@{}ll ll  ll  ll ll ll ll ll r@{\extracolsep{\fill}}l}
\hline\hline
\bf States& $(B^*B)_1$&$(B^{*}_{s}B)_1$&$(B^{*}_{s}B_{s})_{1}$\\
\hline
\bf Parameters&&&\\
$t_{c}$ $[\GeV]^2$&190-230&190-230&190-230\\
$\tau$ $[\GeV]^{-2}\times 10^2$&$~~6~~;~9$&$~~5~~;~9$&$~~7~~;~11$\\

\hline
\hline
\end{tabular*}
}}
\label{tab:b-param}
\end{table*}
\cleardoublepage



\begin{thebibliography}{999}
\bibitem{ANR2} R. M. Albuquerque, S. Narison and D. Rabetiarivony, {\it Phys. Rev.} {\bf D 103}, 074015 (2021).
\bibitem{AFNR} R. Albuquerque et {\it al.}, {\it Phys. Lett.} {\bf B 175}, (2012) 129.
\bibitem{SNX1} R. Albuquerque et {\it al.}, {\it Int. J. Mod. Phys.} {\bf A31} (2016) no.17, 1650093.
\bibitem{SNX2} R. Albuquerque et {\it al.}, {\it Int. J. Mod. Phys.} {\bf A31} (2016) no. 36, 1650196.
\bibitem{SU3} R. Albuquerque et {\it al.}, {\it Int. J. Mod. Phys.} {\bf A33} (2018), 1850082.
\bibitem{ANRR1} R. M. Albuquerque et {\it al.}, {\it Nucl. Part. Phys. Proc.} {\bf 282-284}, (2017) 83.
\bibitem{ANRR1a} R. M. Albuquerque et {\it al.}, {\it Nucl. Part. Phys. Proc.} {\bf 300-302}, (2018) 186-195.
\bibitem{QQQQ} R. M. Albuquerque et {\it al.}, {\it Phys. Rev.} {\bf D 102}, 094001 (2020).
\bibitem{ANRR2} R. M. Albuquerque et {\it al.}, {\it Nucl. Part. Phys. Proc.} {\bf 312-317}, (2021) 120-124.
\bibitem{ANR1} R. Albuquerque et {\it al.}, {\it Nucl. Phys.} {\bf A 1007} (2021) 122113.
\bibitem{ADKT} R. M. Albuquerque et {\it al.}, {\it J. Phys.} {\bf G 46}, 093002 (2019).
\bibitem{BELL} J.S. Bell and R.A. Bertlmann, {\it Nucl. Phys.} {\bf B177} (1981) 218.
\bibitem{BELLa} J.S. Bell and R.A. Bertlmann, {\it Nucl. Phys.} {\bf B187} (1981) 285.
\bibitem{BNR} C. Becchi et {\it al.}, {\it Z. Phys.} {\bf C 8}, 335 (1981).
\bibitem{BERT} R.A. Bertlmann, {\it Acta Phys. Austriaca} {\bf 53}, 305 (1981) and references therein.
\bibitem{NEUF}R.A. Bertlmann and H. Neufeld, {\it Z. Phys.} {\bf C 27} (1985)  437.
\bibitem{SNR}S. Narison and E. de Rafael,  {\it Phys. Lett.} {\bf B 522} (2001) 266.
\bibitem{SVZa} M.A. Shifman, A.I. Vainshtein and V.I. Zakharov, {\it Nucl. Phys.} {\bf B147} (1979) 385, 448.
\bibitem{Za} V.I. Zakharov, {\it Int. J. Mod. Phys.} {\bf A 14}, 4865 (1999).
\bibitem{SNB1} S. Narison, {\it QCD as a theory of hadrons, Cambridge Monogr. Part. Phys. Nucl. Phys. Cosmol.} {\bf 17} (2002) 1; [hep-ph/0205006].
\bibitem{SNB2}S. Narison, {\it QCD spectral sum rules ,  World Sci. Lect. Notes Phys.} {\bf 26} (1989) 1.
\bibitem{SNB3}S. Narison, {\it Phys. Rept.}  {\bf 84} (1982) 263; {\it Acta Phys. Pol.} {\bf B 26} (1995) 687. 
\bibitem{CK} E. de Rafael, hep-ph/9802448.
\bibitem{YND} F. J. Yndurain, {\it The Theory of Quark and Gluon Interactions,} 3rd ed. (Springer, New York, 1999).
\bibitem{PAS} P. Pascual and R. Tarrach, {\it QCD: Renormalization for Practitioner} (Springer, New York, 1985).
\bibitem{RRY} L. J. Reinders, H. Rubinstein, and S. Yazaki, {\it Phys. Rep.} 127, 1(1985).
\bibitem{IOFF} B. L. Ioffe, {\it Prog. Part. Nucl. Phys.} 56, 232(2006).
\bibitem{DOSCH} H. G. Dosch, {\it Non-Perturbative Methods}, edited by S. Narison (World Scientific, Singapor,1985).
\bibitem{BELLE1} Z. Q. Liu et {\it al.} (Belle Collaboration), {\it Phys. Rev. Lett.} 110, 252002 (2013).
\bibitem{BES1} M. Ablikim et {\it al.} (BESIII Collaboration), {\it Phys. Rev. Lett.} 110, 252001 (2013).
\bibitem{BES2} M. Ablikim et {\it al.} (BESIII Collaboration), {\it Phys. Rev. Lett.} 126, 102001 (2021).
\bibitem{LHCb} r. Aaij et {\it al.} (LHCb Collaboration), {\it Phys. Rev. Lett.} 127, 082001 (2021).
\bibitem{PICH}A. Pich and E. de Rafael, {\it Phys. Lett.}  {\bf B158} (1985)  477.
\bibitem{NPIV} S. Narison and A. Pivovarov,  {\it Phys. Lett.} {\bf B 327} (1994) 341.
\bibitem{DJB} D. J. Broadhurst, {\it Phys. Lett.} {\bf 101B}, (1985) 423.
\bibitem{CS} K. G. Chetyrkin and M. Steinhauser, {\it Phys. Lett.}  {\bf B 502}, 104 (2001).
\bibitem{GKPR} P. Gelhausen, A. Khodjamirian, A. A. Pivovarov and D. Rosenthal, {\it Phys. Rev.} {\bf D 88}, 014015 (2013); {\bf 89}, 099901 (E) (2014); {\bf 91}, 099901(E) (2015).
\bibitem{SN18} S. Narison, {\it Int. J. Mod. Phys.} {\bf A 33}, 1850045 (2018); 33, 1850045(A) (2018) and references therein.
\bibitem{SN18a} S. Narison, arXiv:1812.09360.
\bibitem{SN181} S. Narison, arXiv:2101.12578; {\it Nucl. Part. Phys. Proc.} 300-302, 153 (2018).
\bibitem{SN182} S. Narison, {\it Phys. Lett. }{\bf B 784}, 261 (2018).
\bibitem{SN183} S. Narison, {\it Phys. Lett. }{\bf B 802}, 135221 (2020).
\bibitem{SN184} S. Narison, {\it Phys. Lett. }{\bf B 718}, 1321 (2013).
\bibitem{SN185} S. Narison, {\it Int. J. Mod. Phys.}{\bf A 30}, 1550116 (2015); {\it Phys. Lett. }{\bf B 738}, 346 (2014).
\bibitem{ANN} R. M. Albuquerque, S. Narison and M. Nielsen, {\it Phys. Lett.}{\bf B 684}, 236 (2010).
\bibitem{CDKS} Y. Chung, H. G. Dosch, M.Kremer and D. Schall, {\it Z. Phys.} {\bf C 25}, 151 (1984).
\bibitem{BIOFF} B. L. Ioffe, {\it Nucl. Phys.} {\bf B 188}, 317 (1981); {\bf B 191}, 591(E) (1981).
\bibitem{OPF} A. A. Ovchinnikov and A. A. Pivovarov, {\it Yad. Fiz.} {\bf 48}, 1135 (1988).
\bibitem{SN186} S. Narison, {\it Phys. Lett. }{\bf B 605}, 319 (2005).
\bibitem{SNB8} S. Narison, {\it Phys. Lett.}{\bf  B 693}, 559 (2010) ; {\bf 705}, 544(E) (2011).
\bibitem{SNB8a} S. Narison, {\it Phys. Lett.}{\bf  B 693}, 559 (2010) ; {\bf 705}, 544(E) (2011).
\bibitem{SNB8b} S. Narison, {\it Phys. Lett.}{\bf  B 706}, 412 (2011).
\bibitem{SNB8c} S. Narison, {\it Phys. Lett.}{\bf  B 707}, 259 (2012).
\bibitem{LNT} G. Launer, S. Narison and R. Tarrach, {it Z. Phys.}{\bf C 26}, 433 (1984).
\bibitem{SN187} S. Narison, {\it Phys. Lett. }{\bf B 673}, 30 (2009).
\bibitem{DJN} H. G. Dosch, M. Jamin and S. Narison, {\it Phys. Lett. }{\bf B 220} 251 (1989).
\bibitem{BLR}R. A. Bertlmann, G. Launer and E. de Rafael, {\it Nucl. Phys.} {\bf B 250}, 61 (1985).
\bibitem{LMS} W. Lucha, D. Melikhov and H. Sazdjian, {\it Phys. Rev. }{\bf D 100}, 014010 (2019); {\bf 100}, 074029 (2019).
\bibitem{AZIZI} S. S. Agaev, K. Azizi and H. Sundu, {\it  Eur. Phys. J.} {\bf C 77}( 2017) 836.
\end{thebibliography}
\end{document}